**The Pulsar Search Collaboratory: Expanding Nationwide**

Kathryn Williamson, West Virginia University
Maura McLaughlin, West Virginia University
Sue Ann Heatherly, Green Bank Observatory
John Stewart, West Virginia University
Duncan Lorimer, West Virginia University
Harsha Blumer, West Virginia University
Cabot Zabriskie, West Virginia University
Ryan Lynch, Green Bank Observatory

The Pulsar Search Collaboratory (PSC) engages high school students and teachers in analyzing real data from the Robert C. Byrd Green Bank Telescope for the purpose of discovering exotic stars called pulsars. These cosmic clocks can be used as a galactic-scale detector of gravitational waves, ripples in space-time that have recently been directly detected from the mergers of stellar-mass black holes. Through immersing students in an authentic, positive learning environment to build a sense of belonging and competency,[1] the goal of the PSC is to promote students' long-term interests in science and science careers. PSC students have discovered 7 pulsars since the start of the PSC in 2008. Originally targeted at teachers and students in West Virginia, over time the program has grown to 18 states. In a new effort to scale the PSC nationally, the PSC has developed an integrated online training program with both self-guided lectures and homework and real-time interactions with pulsar astronomers. Now, any high school student can join in the exciting search for pulsars and the discovery of a new type of gravitational waves.

**Searching for Pulsars**

Discovered by Jocelyn Bell Burnell in 1967, pulsars have become valuable subjects of research. Pulsars serve as natural laboratories with environments that cannot be replicated on Earth. Made primarily of neutrons, they have extreme densities – more mass than the Sun compressed into the size of a city. They have extreme magnetic fields (on the order of $10^{11}$ T) – over 100 trillion times the strength of a refrigerator magnet, or 10 quadrillion times stronger than Earth's magnetic field.[2] They spin extremely quickly and regularly, some with periods as short as a millisecond.[3] Swept around by this rotational motion, there is a critical distance from the star where the magnetic field lines would have to exceed the speed of light to keep up, and so they break. The radiation created by spiraling electrons is free to escape along the north and south magnetic poles. The radio light beams shine out like a lighthouse, sweeping across the long distances of space, and we detect them here on Earth with our radio telescopes.

The stability of a pulsar's radio beam arrival time at Earth rivals the accuracy of an atomic clock, making pulsars useful timing devices for many fundamental physics applications, including gravitational wave research. If a gravitational wave passes through Earth, it will compress and stretch spacetime so that some pulsars' signals will arrive slightly *earlier* than normal and some will arrive slightly *later*. By looking for correlated changes in pulse arrival times across a network of many pulsars, astronomers in the North American Nanohertz Observatory for Gravitational Waves (NANOGrav)[4,5] and the International Pulsar Timing Array,[5] among others, are poised to detect gravitational waves from the most supermassive black holes at the centers of colliding galaxies billions of lightyears away. These black holes are many millions of times larger and much farther away than the merging black hole pairs recently detected by LIGO,[7] and their discovery will open a whole new window to the gravitational wave universe. To increase the sensitivity of their gravitational wave experiment, NANOGrav must sort through data taken with radio telescopes to discover pulsars to add to the array;[4] therefore, PSC students are part of a large world-wide effort and their work has high scientific value.

Why is searching for pulsars valuable educationally? The positive impact on students of analyzing *real* data with *real* astronomers to make *real* pulsar discoveries is described in Rosen et al.[8] The number one reason students cite for their engagement in the PSC is the ability to work with their friends in a fun learning environment. Other reasons include: the intrinsic enjoyment that comes from thinking about astronomy and pulsars, the chance to visit with others and present their work at the PSC's annual Capstone events on college campuses, praise and encouragement from their teacher, and the feeling of being part of a team that is really contributing to science. Active PSC participants showed significant gains in science identity, STEM career interest, and self-efficacy, with these positive outcomes particularly enhanced for young women. Because these outcomes are generally related to greater persistence in a STEM major in college,[9] the PSC can truly impact students' lives (Figure 1).

> *"You're using the world as a lab, versus some type of simulation or the textbook. And discovering a star is incredible, it's electrifying. I mean, it's like, you discover something that nobody else has ever seen in the universe, and who has an opportunity to do that, you know, in day-to-day life?... I thought it was very exciting, but then as I was doing the data I kind of started figuring out that it's not the discovery that's the most important, it's the process."* – Katya, Teenage Radio Wave Hunter[10]
>
> *"It's kind of what motivated me to, like, actually go into physics. I think without it, without finding that [pulsar], or without even knowing about the program, I wouldn't have gone to college for physics, or even probably science."* - Shay, little green men film[11]

Fig. 1: Quotes from students in the Pulsar Search Collaboratory

**History of the PSC**
In Summer 2008, the Robert C. Byrd Green Bank Telescope (GBT) scanned the sky and archived the radio data to search for the repetitive blips from pulsars. Due to the normal background hum of the telescope receiver noise and data corruption from human-made radio signals, significant data processing is needed to reveal the weak pulsar signals. Data from the GBT are searched for periodic signals using Fourier analysis, which can reveal weak periodic signals in noisy data. Then, the time-series data from each sky position are *folded* at each of the 30 most significant periods returned by the Fourier analysis to produce a diagnostic plot like that shown in Figure 2. These plots require the pattern recognition capabilities of the human eye to discern the tell-tale pulsar-like features. If the signal identified by the Fourier analysis is a pulsar, it will be detectable throughout the observation and over a wide range of radio frequencies. It will also have its peak intensity at a non-zero *dispersion measure*, or DM, which indicates how the pulsar signals were affected by traveling through space and is a proxy for distance. These plots are labeled with important quantities like the pulsar's position in the sky (RA and DEC), the spin period (P), and the significance of the detection (reduced chi-squared). Students must learn about all of these metrics in order to gauge whether a candidate signal is from a pulsar. They also analyze plots created from another type of analysis sensitive to radio bursts from sporadic pulsars called Rotating Radio Transients (RRATs) and enigmatic Fast Radio Bursts (FRB) sources that may not necessarily be periodic.

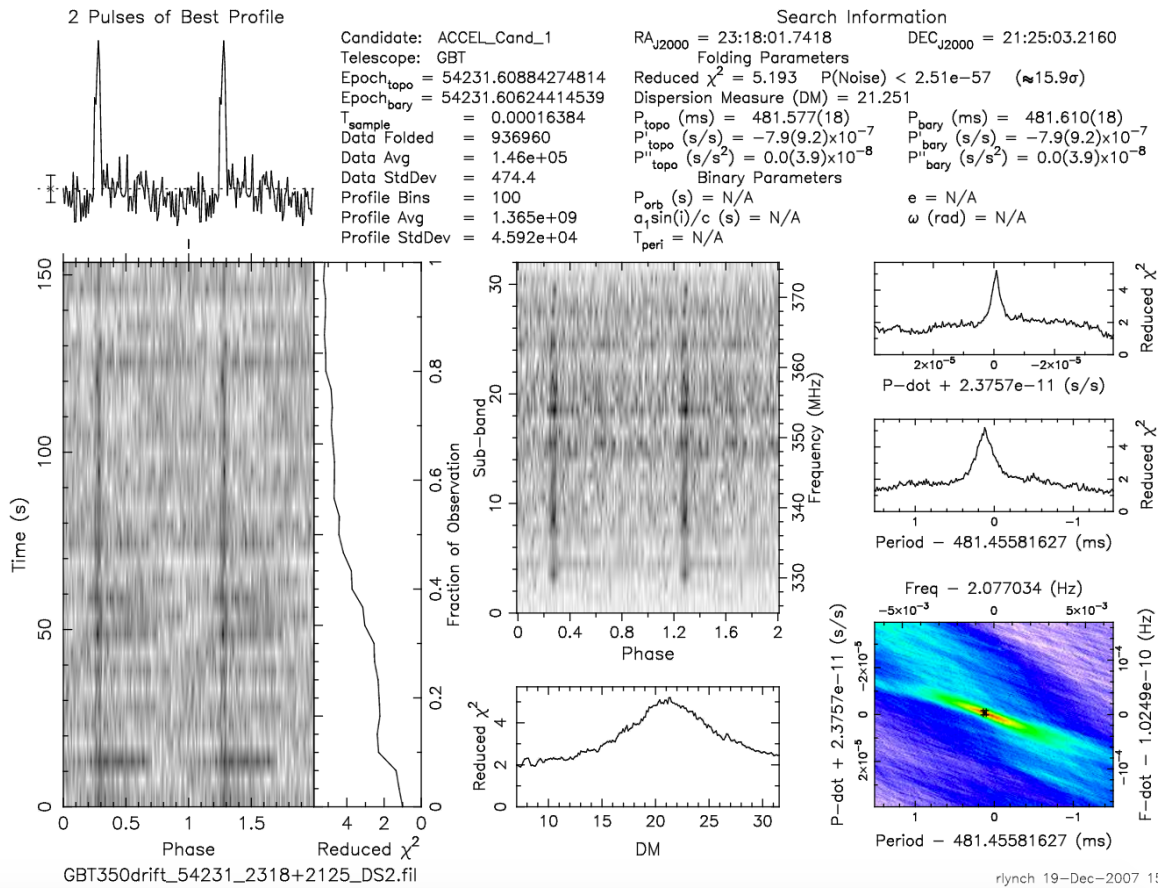

Fig 2: A pulsar data plot, representing the type of plots that PSC students learn to analyze to search for pulsars. This plot shows a pulsar that spins every 0.481 seconds. The tell-tale pulsar features include consistent vertical dark bands on both the left and middle plots, a reduced chi-squared greater than 3, and a non-zero dispersion measure. The plot in the upper left shows the *pulse profile*, or the result of adding every 0.481 seconds of data to all of the previous 0.481 seconds of data. This is a map of the pulsar's emission beam. Two cycles of rotation are shown for clarity.

Hundreds of thousands of plots like that shown in Figure 2 were produced through the observations taken over the summer of 2008. The PSC was then formed through a joint effort between the Green Bank Observatory (GBO) and West Virginia University (WVU) to involve high school teachers and students in the effort to sort through these plots, discover pulsars, and characterize the radio sky. It is important to emphasize that this data is completely "owned" by the students – i.e. no other astronomer has viewed or can view the data. To become trained to reliably analyze the plots, teachers and student leaders attended an in-person, one- to two-week "Summer Institute" at the GBO to learn from professional radio astronomers. They then created clubs at their schools to continue throughout the year and recruit more students. Each new participant was required to pass certification tests with old data to show they could reliably analyze pulsar plots before gaining access to the database of new data to begin their search for pulsars.

For each data plot, PSC students "scored" sub-features of the plot through an online interface to discern whether they thought it was consistent with a pulsar, noise, or human-made radio frequency interference (RFI). To prevent an interesting pulsar candidate from being overlooked, each plot was scored by five students. If a plot was interesting enough, those students were invited to help conduct a follow-up observation with the GBT of that position in the sky. If the follow-up confirmed a pulsar discovery, a refereed paper was published announcing the discovery to the astronomy community, with all students who exceeded some plot inspection threshold listed as co-authors. Since the beginning of the PSC, students have discovered 7 pulsars as well as 1 rotating radio transient, proving that high school students can make significant scientific contributions. Because each plot is inspected by five students, 40 high school students have made a discovery. These discoveries brought tremendous excitement and substantial recognition to those lucky and persistent students.

The excitement is contagious, and the positive outcomes for the thousands of students who have participated in the PSC are significant, even if they were not fortunate enough to discover a pulsar.[8] The PSC celebrated the act of searching itself as an important part of the scientific process, so any student who contributed significant effort (i.e. scored over 2,000 pulsar plots) was invited to attend an annual end-of-year "Capstone" weekend at West Virginia University. At Capstone, students presented posters to showcase their pulsar research, examples of which included investigations into certain individual pulsars, characterizations of noise and RFI, and models of pulsar locations throughout the galaxy. Capstone was also a chance to explore college life, visit research laboratories, meet other scientists, and celebrate the year's hard work with fun team-building activities, leading to some of the most positive experiences cited by students.

**Expanding Nationwide**
While the original PSC required students and teachers to live within driving distance of the GBO and WVU, in 2016 an expansion of the PSC encouraged students from across the nation to join. This expanded version of the PSC has four key features:

(1) A new and growing set of data from the GBT: New data from 375 hours of observations in 2016 are expected to produce 5-10 new discoveries, and as of the writing of this article only about 1% of this data has been examined. With 100 additional observing hours scheduled for late 2018, 2-4 additional discoveries may be possible.
(2) A rich, robust online environment for learning and sharing: With an integrated online training program, visiting GBO for the PSC Summer Institute is no longer the only way to learn how to score pulsar plots. Students and teachers can participate in web-based training synchronously during a series of six one-hour webchats offered each Fall and Spring, or they can follow along asynchronously with pre-recorded video segments.
(3) Additional face-to-face opportunities: The end of year Capstone event is now offered at University of Wisconsin-Madison and Shepherd University in addition to WVU, and support is offered for other universities wishing to host their own Capstone. The yearly PSC Summer Camp at the Green Bank Observatory also now supports particularly active students from anywhere in the nation to come and meet other students like themselves. These face-to-face opportunities are enhanced by a new team of college mentors who help personalize the PSC experience and offer near-peer mentoring on a vast array of topics, such as applying to college and choosing a major.
(4) Advanced research opportunities that allow participants to progress in the community of practice: Participants are now given the opportunity to collect their own radio data by controlling the GBO 20-Meter Telescope online via the Skynet Robotic Telescope Network.[12] Advanced students and mentors can also engage in follow-up timing of newly discovered pulsars, search for multi-wavelength counterpart observations, and conduct additional studies of the raw data (i.e. before the Fourier analysis searching process). These activities develop key data analysis skills that are transferrable well beyond astronomy, helping to prepare students for a broad range of careers that involve "big data."

This paper is an invitation to high school students and teachers across the United States to join in the PSC effort. Students can join as individuals with a verification from a teacher at their school. Teachers who want to engage with students generally do so in one of two ways: either as an advisor for a PSC club (more common) or by incorporating the PSC into an astronomy unit in one of the science classes they teach. Any interested high school student and teacher can get started with the PSC by exploring the website, which includes training materials, manuals, the discussion forum, and information about Summer Institute and Capstone at http://pulsarsearchcollaboratory.com/ and by registering on the database site http://psrsearch.wvu.edu/. To learn more, watch the short Teenage Radio Hunter video[10] or the Little Green Men feature-length movie.[11] Readers may also be interested in similar programs such as the Arecibo Remote Command Center (ARCC)[14] and Pulse@Parkes.[15]

**Acknowledgements**
This work is supported by NSF Advancing Informal STEM Learning (AISL) (awards: #1516512, #1516269). The Green Bank Observatory is a facility of the National Science Foundation operated under a cooperative agreement by Associated Universities, Inc."